\documentclass[
amsmath,
prd,nofootinbib,floatfix,11pt,
]{revtex4}

\newcommand\beq{\begin{eqnarray}}
\newcommand\eeq{\end{eqnarray}}

\def\lsim{\mathrel{\rlap{\lower4pt\hbox{$\sim$}}
    \raise1pt\hbox{$<$}}}                
\def\gsim{\mathrel{\rlap{\lower4pt\hbox{$\sim$}}
    \raise1pt\hbox{$>$}}}

\def\MSbar{$\overline{\rm MS}$ }
\def\lntwo{\ln(2)}
\allowdisplaybreaks
\usepackage{graphicx}
\usepackage{setspace}
\usepackage{axodraw}

\newcommand\Ncolors{N_c}

\begin{document}
\renewcommand{\theequation}{\arabic{section}.\arabic{equation}}

\title{\Large%
QCD corrections to stoponium production at hadron colliders}
\author{James E. Younkin$^1$ and Stephen P. Martin$^{1,2}$}
\affiliation{$^1$Department of Physics, Northern Illinois University, 
DeKalb IL 60115}
\affiliation{$^2$Fermi National Accelerator Laboratory, P.O. Box 500, 
Batavia IL 60510}


\begin{abstract}
If the lighter top squark has no kinematically allowed two-body decays that conserve
flavor, then it will live long enough to form hadronic bound states.
The observation of the diphoton decays of stoponium could then provide a uniquely
precise measurement of the top squark mass.
In this paper, we calculate the cross section for the production of 
stoponium in a hadron collider at next-to-leading order (NLO) in QCD. 
We present numerical results for the cross 
section for production of stoponium at the LHC and study the dependence 
on beam energy, stoponium mass, and the renormalization and factorization scale.
The cross-section is substantially increased by the NLO
corrections, counteracting a corresponding decrease found earlier in the NLO diphoton branching
ratio. 
\end{abstract}

\maketitle
\tableofcontents
\baselineskip=14.9pt

\newpage
\setcounter{footnote}{1}
\setcounter{page}{1}
\setcounter{figure}{0}
\setcounter{table}{0}

\section{Introduction\label{sec:intro}}
\setcounter{footnote}{1}
\setcounter{equation}{0}

In low-energy supersymmetry, the usual collider signatures feature missing energy
carried away by stable, neutral, weakly-interacting particles. Therefore,
there are typically no kinematic mass peaks that could be used to precisely measure
the individual superpartner masses. Other kinematic features can be used to
determine mass differences with some precision, if the decay chains are favorable,
but the overall mass scale of the sparticle spectrum is likely to be much more
difficult to pin down accurately.

One exception to this may occur if new particles can form resonances that
will decay by annihilation into only Standard Model particles with strong
or electromagnetic interactions. A particularly interesting example is
the possibility of stoponium, a 
${}^{2S+1}{L}_J = {}^1S_0$ bound state 
of the lighter top squark and top antisquark, 
which we will denote $\eta_{\tilde t}$.
If the light top squark (or stop) $\tilde t_1$ has no kinematically allowed
flavor-preserving two-body decays, then it will live long enough to form
hadronic bound states including stoponium. 
Specifically, this will occur if the decays
$\tilde t_1 \rightarrow b \tilde C_1$ and $\tilde t_1 \rightarrow t \tilde N_1$
are kinematically forbidden, where $\tilde C_1$ and $\tilde N_1$ are the
lighter chargino and the lightest neutralino respectively.
As pointed out originally by Drees and Nojiri 
\cite{Drees:1993yr,Drees:1993uw},
and re-examined recently in \cite{Martin:2008sv,Martin:2009dj}, 
decays of $s$-wave stoponium to $\gamma\gamma$ can provide a viable signal
at the Large Hadron Collider (LHC) with sufficient beam energy and integrated luminosity.
The signal will appear 
as a peak with width determined by detector resolution on top of a smoothly falling background. This could provide
a uniquely precise measurement of the top squark mass, which would in turn
serve as a ``standard candle" for kinematics of the supersymmetric sector.
(See also refs.~\cite{Nappi:1981ft}-\cite{Kats:2009bv}
for other work related to stoponium at colliders.)

As reviewed in more detail in refs.~\cite{Martin:2008sv,Martin:2009dj},
there are at least two good motivations for considering top squarks light enough
to form bound states. First, in models of ``compressed supersymmetry" \cite{compressed}, the
top squark automatically comes out relatively light (compared to so-called mSUGRA models). These models can provide the right amount of dark matter provided that
the mass difference between $\tilde t_1$ and $\tilde N_1$ is less than the top quark mass, with $m_{\tilde t_1}$ between about 200 GeV and 400 GeV.
Second, minimal supersymmetric models that can provide baryogenesis at the electroweak scale \cite{EWbaryogenesis,EWbaryogenesisnew} require
a quasi-stable stop with a mass that is currently estimated to be
in the range from approximately 118 GeV to 135 GeV.

The production and decay of stoponium as studied in 
refs.~\cite{Drees:1993yr,Drees:1993uw,Martin:2008sv} at leading order (LO) are subject to large QCD
radiative corrections and a strong dependence on the renormalization
scale. The purpose of the present paper is to compute the corrections to
stoponium production in hadronic collisions at next-to-leading order (NLO)
in QCD, in order to help interpret limits on, or discovery of, 
a stoponium resonance at the LHC. This result will complement (and will make use of) our previous
calculation of the QCD NLO corrections to stoponium decay \cite{Martin:2009dj}.

In this paper, we will model the inclusive production of stoponium at a hadron collider as factored into the production of the free 
squark-antisquark pair with the same momentum in a color-singlet state and the binding of the pair into the stoponium bound state.  For each 
pair of initial-state partons, the production amplitude for the squark-antisquark state with quantum numbers identical to the scalar ground state of stoponium is 
calculated in perturbative QCD, then related to the parton-level 
stoponium production cross section by integrating over the phase space 
that models the non-perturbative process of bound state formation using the 
wavefunction at the origin.
In carrying out this program, we will closely follow the methods of 
ref.~\cite{Kuhn:1992qw}, which treated the analogous case of ${}^1S_0$ toponium
($\eta_t$) production. (Ref.~\cite{Kuhn:1992qw} was written before the mass of 
the top quark was known to be too large to allow it to form bound states, so
toponium was a viable possibility at that time.)

It is well known that the ``static color singlet model" just described has 
failed spectacularly in explaining the observed
$J/\psi$ and $\psi(2S)$ prompt production in hadronic collisions. 
The color singlet model prediction for the prompt 
production cross section of the $\psi(2S)$ charmonium 
state is too small by a factor of about 50, while the prediction
for prompt $J/\psi$ is too small by a factor of about 6, compared to the Tevatron data
\cite{Abe:1997jz}. 
(The $\Upsilon$ production cross section data \cite{UpsilonCDF,UpsilonDzero}
seems to be acceptably fit by the color singlet model 
after inclusion of NNLO contributions
\cite{UpsilonNNLO}.) 
The failure of the static color singlet model in the charmonium case 
is due in part to the assumption that the hadronization of 
$Q\overline{Q}$ can be factored into perturbative calculations 
of the production of open $Q\overline{Q}$ with color and spin identical to the bound state and the nonrelativistic probability for annihilation at the origin. 
The discrepancy has provoked a variety of efforts to go beyond the static color singlet model in calculations; for reviews see 
\cite{Petrelli:1997ge}-\cite{Lansberg:2008gk}.
For example,
more general analyses performed using the 
NRQCD effective theory for nonrelativistic heavy quarks 
show that large enhancements to quarkonium production can 
come from the production of the constituent quarks in several different color and 
angular momentum states that can transition to the desired bound state. In this 
approach, the parton-level differential cross section for the 
production of a bound state $\Phi$ in a collider 
is given by:
\beq
d\hat{\sigma}(ab \rightarrow \Phi + X) = \sum_n d\hat{\sigma}(ab \rightarrow Q\overline{Q}[n]+X) \langle \mathcal{O}^\Phi [n] \rangle,
\label{eq:beyondCSmodel}
\eeq
Here the differential cross section is factored into the short-distance perturbatively calculated cross sections initiated by partons $a,b$ for the 
production of a quark-antiquark pair $Q\overline{Q}$ 
in the color and spin state $n$, and the long-distance 
transition probabilities $\langle \mathcal{O}^\Phi[n] \rangle$ that describe the nonperturbative component of the hadronization. 
The sum over states $n$ can be thought of as a perturbative series in the relative velocity $v$, which separates the long- and short-distance scales.  So-called power counting rules are used to determine the relative order of each transition based on the 
effective theory interactions it represents \cite{Lepage:1992tx,Bodwin:1994jh}.  
The series converges more quickly as the mass of the quarks 
increases; 
for charmonium $v$ is approximately 0.3 and for bottomonium $v \sim 0.1$.  
Transitions suppressed by 
the relative velocity can nevertheless have large cross sections 
and be much larger than the direct production assumed in the color singlet model, 
which may help explain the large observed prompt production cross sections.
However, measurements of the $J/\psi$ polarization \cite{Psipolarization}
at the Tevatron and other 
measurements \cite{PHENIX} by PHENIX at RHIC do not seem to fit well the predictions
of the NRQCD approach.  

Fortunately, however, the importance of corrections
beyond the static color-singlet model should be relatively much less important for 
stoponium production, for several reasons. First, because the top squarks are much heavier, expansions in $\alpha_S$ should converge more quickly.
Second, because $\eta_{\tilde t}$ formed from scalars
is a $J=0$ state, the LO cross section for the color singlet state does not require an extra gluon, as is the case for $J/\psi$, $\psi(2S)$ or $\Upsilon$ production at leading order. 
Finally, in the stoponium case, all color octet states that can transition to the color singlet stoponium state ${}^1S_0$ are of relative order $v^4$.  They are already negligible in the case of $\eta_b$ production at next-to-leading order
\cite{Maltoni:2004hv}, and since $v$ scales with $\alpha_S$,
these color-octet transitions should be suppressed by an additional large factor
for stoponium.  
These considerations justify our use of the color singlet model 
to calculate the radiative corrections to scalar stoponium 
production and decay in perturbative QCD.
Note that this essentially amounts to letting the wavefunction(s) at the origin
$R(0)$ in a 
potential model play the role of the spin-0 color-singlet matrix element(s) in eq.~(\ref{eq:beyondCSmodel}), 
while neglecting the effects of 
the higher spin or color matrix elements. A more accurate treatment 
including those effects may eventually be necessary, 
but is beyond the scope of the present paper and in
any case would require some way of 
estimating the other relevant matrix elements,
which are presently unavailable.

The organization of the remainder of this paper is as follows.
In section II we provide the NLO parton-level cross sections 
for stoponium production in hadronic collisions in the static 
color singlet model. In section III we discuss how to turn these calculations
into hadron-collision cross sections.
In section IV, we provide numerical results for the NLO stoponium production cross
section in proton-proton collisions, studying
the dependence on beam energies relevant for the CERN Large Hadron Collider (LHC),
on the stoponium mass, and on the renormalization and factorization scale.
We also review the next-to-leading order hadronic and diphoton branching ratios for the annihilation decays found in 
\cite{Martin:2009dj}, 
which we use to estimate the NLO QCD 
cross section times branching ratio for the observable signal
$pp \rightarrow \eta_{\tilde t} +X \rightarrow \gamma \gamma +X$ 
in scenarios where the hadronic partial width dominates the full width,
and discuss corrections that apply to this idealized result
in compressed supersymmetry and in models with electroweak-scale baryogenesis.
   
\section{Parton-level cross sections\label{sec:parton}}
\setcounter{footnote}{1}
\setcounter{equation}{0}

In this section, we calculate the individual parton-level cross sections
that contribute to the hadronic production of stoponium in NLO QCD.  We
will closely follow the method used by K\"uhn and Mirkes to calculate
radiative corrections to toponium production in ref.~\cite{Kuhn:1992qw}.  Although
the virtual corrections to the leading-order diagrams for stoponium differ
from the toponium result, we will show that the other radiative
corrections contributing to the next-to-leading order cross section have
the same form for stoponium as for their toponium counterparts, when both
are written in terms of their corresponding leading order results. The
relevant Feynman rules for our calculation can be found in
\cite{Martin:2009dj}. Both ultraviolet and infrared divergences will
be dealt with by dimensional regularization.

In order to compute the parton-level cross-sections for stoponium
production from the rate for open squark-antisquark production,
integration over the final-state phase space must be restricted to the
subspace containing the bound state.  Let $d{\mathcal{P}}_N
(k_1,k_2;p_1,p_2,\ldots,p_N)$ be the Lorentz-invariant differential phase
space element for an $N$-body final state.  In our case, $N=2$ or 3, and
$k_1$ and $k_2$ will be the initial (massless) parton momenta and $p_1$
and $p_2$ are the final squark and antisquark momenta and $p_3$ is a
possible final state (massless) parton momentum. To project onto the bound
state phase space, we require the final-state squarks to have identical
4-momenta, letting $p = p_1 = p_2$, so that $2p$ is the stoponium
momentum, and use the radial wavefunction at the origin $R(0)$ 
(normalized so that $\psi(\vec{r}) = R(r)/\sqrt{4\pi}$ for an $S$-wave
state) to
characterize the long-range behavior of the squark hadronization.  As a
consequence of this projection, the relevant 2-body and 3-body
differential phase space factors in $d = 4 - 2 \epsilon$ dimensions are
replaced by
\beq
d\mathcal{P}_2 & \rightarrow & \frac{|R(0)|^2}{M} \delta(\hat{s}-M^2) 
\label{eq:dP2},
\\
d\mathcal{P}_3 & \rightarrow & 
\frac{|R(0)|^2}{16\pi^2 \Gamma(1-\epsilon) M } \left( 
\frac{4\pi \hat{s}}{\hat{u} \hat{t}} \right)^{\epsilon} 
\frac{d\hat{t}}{\hat s},
\label{eq:dP3}
\eeq
where the parton-level Mandelstam variables are defined by
\beq
\hat{s} = (k_1 + k_2)^2,
\qquad
\hat{t} = (2p - k_1)^2,
\qquad
\hat{u} = (2p - k_2)^2,
\label{eq:definestu}
\eeq
and satisfy
\beq
\hat{s} + \hat{t} + \hat{u} = 4 p^2 = M^2 
\eeq
where $M = 2 m_{\tilde t} - E_{\rm binding} \approx 2 m_{\tilde t}$ is the bound state mass.
Here and in the remainder of the paper, we will often write $\tilde t$ without
the subscript 1 and $M$ instead of $M_{\eta_{\tilde t}}$, for simplicity.

\subsection{Gluon fusion at leading order}

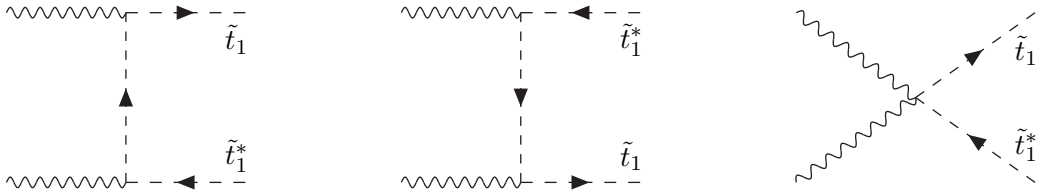
\begin{figure}[!tbp]
\begin{center}
\begin{picture}(80,80)(-40,-40)
\Photon(-45,-32)(0,-32){2}{8}
\Photon(-45,32)(0,32){2}{8}
\DashArrowLine(45,-32)(0,-32){4}
\DashArrowLine(0,-32)(0,32){4}
\DashArrowLine(0,32)(45,32){4}
\Text(42,22)[]{$\tilde{t}_1$}
\Text(42,-22)[]{$\tilde{t}_1^*$}
\end{picture}
~~~~~~~~~~~~~~~~~
\begin{picture}(80,80)(-40,-40)
\Photon(-45,-32)(0,-32){2}{8}
\Photon(-45,32)(0,32){2}{8}
\DashArrowLine(0,-32)(45,-32){4}
\DashArrowLine(0,32)(0,-32){4}
\DashArrowLine(45,32)(0,32){4}
\Text(42,22)[]{$\tilde{t}_1^*$}
\Text(42,-22)[]{$\tilde{t}_1$}
\end{picture}
~~~~~~~~~~~~~~~~~
\begin{picture}(80,80)(-40,-40)
\Photon(-45,-32)(0,0){2}{8}
\Photon(-45,32)(0,0){2}{8}
\DashArrowLine(45,-32)(0,0){4}
\DashArrowLine(0,0)(45,32){4}
\Text(42,18)[]{$\tilde{t}_1$}
\Text(42,-18)[]{$\tilde{t}_1^*$}
\end{picture}
\end{center}
\caption{\label{fig:leadingorder}
The leading-order diagrams for the production of stoponium.}
\end{figure}
The parton-level diagrams for the leading order production of stoponium
are in Fig.\ \ref{fig:leadingorder}.  In this calculation, the squark and
antisquark are restricted to color combinations appropriate for the
singlet bound state using a wavefunction projection factor
$\delta_i^j/\sqrt{\Ncolors}$, where $i,j$ are $SU(\Ncolors)$ fundamental
representation indices (with $\Ncolors = 3$ in the real world). The
squared amplitude obtained from these diagrams, averaged over initial
color and polarization and summed over final color, is
\beq
\frac{1}{(d-2)^2(\Ncolors^2-1)^2} \sum{|\mathcal{M}|^2} = 
\frac{8 \pi^2 \widehat\alpha_S^2 
\mu^{4\epsilon}}{\Ncolors (\Ncolors^2 - 1)(1-\epsilon)},
\eeq
where $\mu$ is the regularization scale and $\widehat\alpha_S$ is the bare
coupling. 

The leading order differential cross section 
for stop-antistop production in a color singlet state is therefore
\beq
d\hat\sigma_{\rm LO}(gg\rightarrow \tilde t \tilde t^*) 
= \frac{1}{2\hat{s}} \frac{1}{(d-2)^2(\Ncolors^2-1)^2} \sum{|\mathcal{M}|^2} 
d\mathcal{P}_2,
\eeq
where $\hat{s}$ is the parton-level center-of-momentum energy squared.
The leading-order cross-section for scalar stoponium 
production is obtained from this by the projection that restricts the
squark and antisquark to identical
4-momenta and includes the probability of annihilation at the origin,
using eq.~(\ref{eq:dP2}):
\beq
\hat\sigma_{\rm LO}(gg\rightarrow \eta_{\tilde{t}}) =  
\frac{4 \pi^2 \widehat \alpha_S^2 \mu^{4 \epsilon}}{
\Ncolors (\Ncolors^2 - 1) (1 - \epsilon)} \frac{|R(0)|^2}{M^3} 
\frac{\delta(1-z)}{\hat s},
\label{eq:LOcross}
\eeq
where
\beq
z \equiv M^2/\hat s.
\label{eq:definez}
\eeq
Note that the result (\ref{eq:LOcross}) for LO stoponium production is a
factor of $2(1 - 2 \epsilon)$ smaller than the corresponding result for
toponium obtained in eq.~(10) of ref.~\cite{Kuhn:1992qw}. 

\subsection{Gluon fusion at NLO}

The corrections to leading-order gluon fusion can be divided into two 
parts - the virtual corrections coming from gluon loops in the 
leading-order diagrams and the corrections that involve the real emission 
of an additional gluon in the final state. The virtual corrections to 
stoponium production through gluon fusion are identical to the virtual 
corrections to stoponium annihilation to two gluons, which we have 
already calculated \cite{Martin:2009dj}.  Adding up the two-particle cuts 
in Tables I and II of ref.~\cite{Martin:2009dj}, and combining with 
eq.~(\ref{eq:LOcross}), we find
\beq
{\hat\sigma}_{\mathrm{LO + virtual}} (gg \rightarrow \eta_{\tilde{t}}) & = & 
\sigma_0 \delta(1-z) \frac{\mu^{4 \epsilon}}{1 - \epsilon} \Bigg \{ 1 + \frac{\alpha_S}{\pi} 
\left( \frac{4 \pi \mu^2}{M^2} \right)^{\epsilon} 
\Gamma(1+\epsilon) \Big[ \frac{b_0}{2} \ln \left( \frac{Q^2}{M^2} \right) 
\nonumber \\ && 
- \frac{\Ncolors}{\epsilon^2_{\rm IR}} 
- \frac{b_0}{2\epsilon_{\rm IR}} + \left( 1 + \frac{5 \pi^2}{12} \right) \Ncolors 
- \left( 3 + \frac{\pi^2}{4} \right) C_F \Big] \Bigg \}
,
\label{eq:sigmaLOplusvirtual}
\eeq
where $C_F = (\Ncolors^2 - 1)/2\Ncolors$, and
\beq
b_0 & = & \frac{11}{3} \Ncolors - \frac{2}{3} n_f,
\eeq
with $n_f$ the number of
quark flavors, and we define for future convenience
\beq
\sigma_0 \equiv
\frac{4 \pi^2 \alpha_S^2}{
\Ncolors (\Ncolors^2 - 1) \hat s} \frac{|R(0)|^2}{M^3} .
\label{eq:definesigma0}
\eeq
In eqs.~(\ref{eq:sigmaLOplusvirtual}) and (\ref{eq:definesigma0}), 
$\alpha_S(Q)$ is the \MSbar coupling renormalized at the scale $Q$, 
related at one-loop order to the
bare coupling $\widehat\alpha_S$ and the regularization scale $\mu$ by
\beq
\widehat\alpha_S = \alpha_S \left [1 - \frac{\alpha_S}{\pi} \frac{b_0}{4}
\left ( \frac{1}{\epsilon_{\rm UV}} + 
\ln(4\pi \mu^2/Q^2) - \gamma_E \right ) \right ].
\eeq
Note that we do not include the two-cut diagrams relating to the 
insertion of squark loops in the gluon propagator (diagrams q1-q4 of 
ref.~\cite{Martin:2009dj}) since we will use $\alpha_S^{(n_f)}$ from 
the $n_f=5$-quark effective theory to be consistent with the 
parton distribution functions we will use. We have also made the 
replacement $\ln(m_f^2/M^2) \rightarrow 1/\epsilon_{\rm IR}$ in the mass 
singularity arising from quark loops and the $1/v$ 
singularity is absorbed into the definition of the bound state wavefunction.

\begin{figure}[!tbp]
\begin{center}
\begin{picture}(80,80)(-40,-40)
\Photon(-45,-32)(0,-32){2}{8}
\Photon(-45,0)(0,0){2}{8}
\Photon(-45,32)(0,32){2}{8}
\DashArrowLine(45,-32)(0,-32){4}
\DashArrowLine(0,-32)(0,0){4}
\DashArrowLine(0,0)(0,32){4}
\DashArrowLine(0,32)(45,32){4}
\Text(-52,32)[]{$i$}
\Text(-52,0)[]{$j$}
\Text(-52,-32)[]{$n$}
\Text(42,24)[]{$\tilde{t}$}
\Text(42,-22)[]{$\tilde{t}^*$}
\Text(0,-46)[c]{(a)}
\end{picture}
~~~~~~~~~~~~~~~~~
\begin{picture}(80,80)(-40,-40)
\Photon(-45,-32)(0,-32){2}{8}
\Photon(-45,0)(-22,32){2}{7}
\Photon(-45,32)(0,32){2}{8}
\DashArrowLine(45,-32)(0,-32){4}
\DashArrowLine(0,-32)(0,32){4}
\DashArrowLine(0,32)(45,32){4}
\Text(-52,32)[]{$i$}
\Text(-52,0)[]{$j$}
\Text(-52,-32)[]{$n$}
\Text(42,24)[]{$\tilde{t}$}
\Text(42,-22)[]{$\tilde{t}^*$}
\Text(0,-46)[c]{(b)}
\end{picture}
~~~~~~~~~~~~~~~~~
\begin{picture}(80,80)(-40,-40)
\Photon(-45,-32)(0,0){2}{8}
\Photon(-45,0)(-22,16){2}{5.25}
\Photon(-45,32)(0,0){2}{8}
\DashArrowLine(45,-32)(0,0){4}
\DashArrowLine(0,0)(45,32){4}
\Text(-52,32)[]{$i$}
\Text(-52,0)[]{$j$}
\Text(-52,-32)[]{$n$}
\Text(42,20)[]{$\tilde{t}$}
\Text(42,-18)[]{$\tilde{t}^*$}
\Text(0,-46)[c]{(c)}
\end{picture}
\end{center}
\caption{\label{fig:realgluon}
Non-zero diagrams related by crossing symmetry to the real gluon emission 
corrections to scalar stoponium production.  Diagrams related to (a) and (b) 
by arrow reversal are also included with them but not shown.
}
\end{figure}
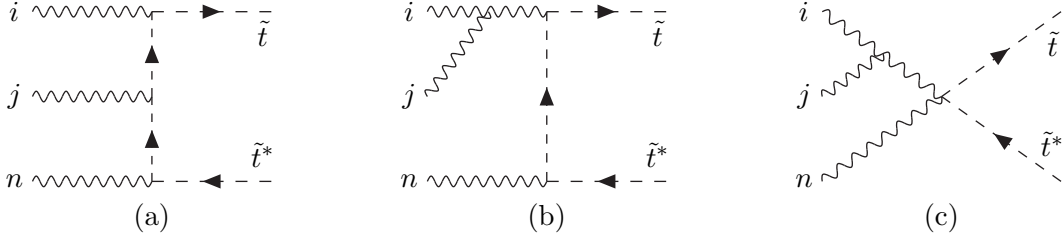

The squared matrix element for real gluon emission can be obtained from 
that of the diagrams for $ggg \rightarrow \tilde t \tilde t^*$ 
in Figure \ref{fig:realgluon} by crossing one gluon to 
the final state in all possible ways.
The three external gluons carry labels $i,j,n = 1,2,3$ and carry
momenta and polarizations $k_i$ and $\epsilon_i$, etc.
The stop and antistop each with mass $m$ have the same 
momentum $p$ (since we let the relative velocity go to zero).
The matrix element resulting from Figure \ref{fig:realgluon} 
is
\beq
\mathcal{M}_{(i,j,n)} = \frac{g^3 f^{a_i a_j a_n}}{\sqrt{\Ncolors}}
\epsilon_i^\mu \epsilon_j^\nu \epsilon_n^\rho
\left [
\mathcal{M}^{(a)}_{\mu\nu\rho} + 
\mathcal{M}^{(b)}_{\mu\nu\rho} + 
\mathcal{M}^{(c)}_{\mu\nu\rho}  
\right ],
\eeq
where
\beq
i \mathcal{M}^{(a)}_{\mu\nu\rho} &=& 
\frac{(k_j + k_n)_\mu (k_n - k_i)_\nu (k_i + k_j)_\rho}{
2[(p-k_i)^2 - m^2][(p-k_n)^2 - m^2]}
,
\\
i \mathcal{M}^{(b)}_{\mu\nu\rho} &=& 
\frac{[ g_{\mu\nu} (k_i - k_j)\cdot k_n
+ (k_i + 2 k_j)_\mu (k_n)_\nu
- (2k_i + k_j)_\nu (k_n)_\mu]
(k_i + k_j)_\rho }{
[(p-k_n)^2 - m^2] (k_i + k_j)^2} 
,
\\
i \mathcal{M}^{(c)}_{\mu\nu\rho} &=& 
[ 
g_{\mu\nu} (k_i - k_j)_\rho + 
g_{\nu\rho} (k_i + 2 k_j)_\mu 
- g_{\rho\mu} (2k_i + k_j)_\nu]
/(k_i + k_j)^2
\eeq
correspond to the contributions from diagrams of types
\ref{fig:realgluon}(a), (b), and (c) respectively.
The total matrix element that includes all diagrams is 
\beq
\mathcal{M} = \sum_{(i,j,n)} \mathcal{M}_{(i,j,n)},
\eeq
where $\sum_{(i,j,n)}$ represents a sum over the six permutations
$(i,j,n) = (1,2,3)$, $(1,3,2)$, $(2,1,3)$, $(2,3,1)$, $(3,1,2)$, $(3,2,1)$.
Taking the labels 1,2 to refer to the initial-state gluons and 3 for the final-state
gluon, the spin-summed squared matrix element for 
$gg \rightarrow \tilde t \tilde t^* g$ is equal to $\sum|\mathcal{M}|^2$ by crossing,
when expressed in terms of the parton-level Mandelstam variables defined in eq.~(\ref{eq:definestu}).

The sums over the physical polarizations of the gluons are performed using
\beq
\sum_\lambda 
\epsilon_{\mu}(\lambda,k) \epsilon_{\nu}^*(\lambda,k) 
= -g_{\mu \nu} + \frac{n_{\mu} k_{\nu} + n_{\nu} k_{\mu}}{n \cdot k} - \frac{n^2 k_{\mu} k_{\nu}}{(n \cdot k)^2} 
\eeq
where $k$ is the momentum of the gluon in question 
and $n$ is an arbitrary 4-vector.  This avoids the added complication 
of needing to cancel unphysical polarizations using additional diagrams 
that include ghost loops for each triple gluon vertex.  Note that it is 
convenient for any given gluon polarization sum 
to choose any other of the massless gluon momenta for $n$. 
Including factors for the averaging over initial gluon polarization and 
color, this parton-level differential cross section is
\beq
d\hat{\sigma} (gg \rightarrow \eta_{\tilde{t}}g) = 
\frac{1}{2\hat{s}} \frac{1}{(d-2)^2(\Ncolors^2-1)^2} \sum |\mathcal{M}|^2 \ 
d\mathcal{P}_3 ,
\eeq
where $d\mathcal{P}_3$ is understood to be replaced according to
eq.~(\ref{eq:dP3}) to obtain the stoponium + X differential phase space.
The result is
\beq
d \hat{\sigma} (gg \rightarrow \eta_{\tilde{t}}g) & = & 
\sigma_0 \frac{\alpha_S}{2\pi} 
\Ncolors \frac{M^2}{\hat{s}} 
\left( \frac{4 \pi \mu^2 \hat{s}}{\hat{u} \hat{t}} \right)^{\epsilon} 
\frac{\mu^{4 \epsilon}}{\Gamma (2-\epsilon)} \Bigg[ 
\frac{\hat{s} \hat{t}+\hat{t} \hat{u}+ \hat{u} \hat{s}}{
(\hat{s}-M^2)(\hat{t}-M^2)(\hat{u}-M^2)} \Bigg]^2 
\nonumber \\ && 
\Bigg[ \frac{M^8 + \hat{s}^4 + \hat{t}^4 + \hat{u}^4}{
\hat{s} \hat{t} \hat{u}} \Bigg] \ d\hat{t}
,
\eeq
where we have dropped a term proportional to $\epsilon$ that does not 
have a potentially singular denominator after angular integration
in the $\epsilon \rightarrow 0$ limit. 

Angular integration is carried out by replacing the Mandelstam variables 
with the dimensionless variables $y$ and $z$ defined on the interval 
$[0,1]$ by eq.~(\ref{eq:definez}) and
\beq
y = (1 + \cos \theta )/2 ,
\eeq 
where $\theta$ is the angle in the center-of-momentum frame between the 
initial-state parton with momentum $k_1$ and the stoponium momentum 
direction. 
This implies that
\beq
\hat{t} = -M^2 (1-y) (1-z)/z 
,
\qquad \qquad 
\hat{u} = -M^2 y (1-z)/z 
.
\label{eq:substitution}
\eeq
In terms of $y$ and $z$, the partonic cross section is 
\beq
\hat{\sigma} (gg \rightarrow \eta_{\tilde{t}}g) &=&  
\sigma_0 \frac{\alpha_S}{\pi} \Ncolors 
\left( \frac{4 \pi \mu^2}{M^2} \right)^{\epsilon} 
\frac{\mu^{4 \epsilon}}{\Gamma (2-\epsilon)} z^{1+\epsilon} (1-z)^{-1-2\epsilon}
\int_0^1 dy \,
y^{-\epsilon} (1-y)^{-1-\epsilon} \nonumber \\ && 
\left( \frac{1 - y(1-y)(1-z)}{[1-y(1-z)][z+y(1-z)]} \right)^2 
\left( z^4 + 1 + (1-z)^4 \left[ (1-y)^4 + y^4 \right] \right) 
,
\label{eq:sigmaggyz}
\eeq
where we have used the symmetry of the rest of the integrand
under $y \rightarrow 1-y$
to replace $y^{-1-\epsilon} (1-y)^{-1-\epsilon}$ with 
$2 y^{-\epsilon} (1-y)^{-1-\epsilon}$. 

To compute this integral, one uses plus distributions to simplify the 
integrand and isolate the soft and collinear 
divergences before integration is carried out. 
The plus distribution $\left( F(x) \right) _+$ of a function $F(x)$ is 
defined by \cite{Field:1989uq}
\beq
\left( F(x) \right) _+ \equiv \lim_{\beta \rightarrow 0} 
\Bigg[ F(x) \Theta( 1 - x - \beta ) - 
\delta(1 - x - \beta) \int_0^{1-\beta} F(y) \, dy \Bigg],
\label{eq:plusdist}
\eeq
from which follows the identity 
\beq
\int_0^1 G(x) \left( F(x) \right){_+} \, dx 
= \int_0^1 \Big[ G(x) - G(1) \Big] F(x) \, dx
\label{eq:1identity}
\eeq
and the expansion identities
\beq
y^{-\epsilon} (1-y)^{-1-\epsilon} = -\frac{1}{\epsilon} \delta (1-y) + 
\left( \frac{1}{1-y} \right) _+ - \epsilon \ \frac{\ln y}{1-y} - \epsilon \ 
\left( \frac{\ln (1-y)}{1-y} \right) _+,
\label{eq:yplus}
\\
z^{\epsilon} (1-z)^{-1-2\epsilon} = 
-\frac{1}{2\epsilon} \delta (1-z) + \left( \frac{1}{1-z} \right)_+ 
+ \epsilon \ \frac{\ln z}{1-z} - 
2 \epsilon \ \left( \frac{\ln (1-z)}{1-z} \right)_+  
.
\label{eq:zplus}
\eeq
Using these expansions in eq.~(\ref{eq:sigmaggyz}) and integrating over $y$ using
eq.~(\ref{eq:1identity}), one obtains:
\beq
\hat{\sigma} (gg \rightarrow \eta_{\tilde{t}}g) &=&  
\sigma_0 \frac{\alpha_S}{\pi} 2 \Ncolors \left( 
\frac{4 \pi \mu^2}{M^2} \right)^{\epsilon} 
\frac{\mu^{4 \epsilon} \Gamma(1+\epsilon)}{1 - \epsilon} 
\Bigg \{ 
\frac{1}{2 \epsilon^2} \delta(1-z) - \frac{\pi^2}{6} \delta(1-z) 
\nonumber \\ && 
- \frac{1}{\epsilon} \frac{[1-z(1-z)]^2}{z}  \left( 
\frac{1}{1-z} \right)_+ - \Big[ \frac{11z^6 + 2z^4 + 24z^3 + 23z^2 + 
12}{12z(1+z)^2} 
\nonumber \\ && 
+ 
\frac{2z^7+3z^6+z^4+2z^3+5z^2-1}{z(1-z)(1+z)^3} \ln z \Big] \left( 
\frac{1}{1-z} \right)_+ 
\nonumber \\ && 
+ \frac{[1 - z(1-z)]^2}{z(1-z)} 
\Big[ 2 (1-z) \left( \frac{\ln (1-z)}{1-z} \right)_+ - \ln z \Big] \Bigg \}.
\eeq
Further simplification can be carried out by noting that if 
$G(1)=0$, then eq.~(\ref{eq:1identity}) implies 
\beq
\int_0^1 G(z) \left( F(z) \right)_+ \, dz  = \int_0^1 G(z) F(z) \, dx,
\label{eq:plussimplify}
\eeq
and therefore one can simply replace the plus distributions multiplying such functions accordingly,
under the assumption that we will eventually integrate to an upper limit 
$z=1$. The result of this simplification is
\beq
\hat{\sigma} (gg \rightarrow \eta_{\tilde{t}}g) &=&  
\sigma_0 \frac{\alpha_S}{\pi} \Ncolors 
\left( \frac{4 \pi \mu^2}{M^2} \right)^{\epsilon} 
\frac{\mu^{4 \epsilon} \Gamma(1+\epsilon)}{1 - \epsilon} 
\Bigg \{ \frac{1}{\epsilon^2} \delta(1-z) - \frac{\pi^2}{3} \delta(1-z) \nonumber \\ && - \frac{2}{\epsilon} \Big[ \frac{1}{z} + \left( \frac{1}{1-z} \right)_+ + z (1-z) -2 \Big] + F_{gg}(z) \Bigg \},
\label{eq:sigmaggreal}
\eeq
where  
\beq
F_{gg}(z) &=& 
\frac{11z^5+11z^4+13z^3+19z^2+6z-12}{6z(1+z)^2} 
+ 4 \left(\frac{1}{z} + z(1-z) - 2 \right) \ln (1-z) 
\nonumber \\ && 
+ 4 \left( \frac{\ln (1-z)}{1-z} \right)_+ 
+ \frac{2(z^3-2z^2-3z-2)(z^3-z+2)z \ln z}{(1+z)^3(1-z)^2} - \frac{3}{1-z}.
\label{eq:defineFgg}
\eeq
This is identical in form to the corresponding real emission correction 
to the leading order parton-level toponium cross section 
\cite{Kuhn:1992qw}, when both are written in terms of their respective 
LO results $\sigma_0$.

To get the full next-to-leading order QCD cross section for gluon fusion, one 
must add the leading order plus virtual corrections from 
eq.~(\ref{eq:sigmaLOplusvirtual}) 
and real emission corrections from (\ref{eq:sigmaggreal}), resulting in
\beq
\hat{\sigma} (gg \rightarrow \eta_{\tilde{t}}+X) &=& 
\sigma_0 \frac{\mu^{4 \epsilon}}{1-\epsilon}
\Bigg \{ \delta(1-z) + \frac{\alpha_S}{\pi} \Gamma (1+\epsilon) 
\left( \frac{4 \pi \mu^2}{M^2} \right)^{\epsilon} 
\Bigg[ \delta (1-z) \bigg \lbrace \frac{b_0}{2} \ln \left( \frac{Q^2}{M^2} \right)  
\nonumber \\ && 
+ (\Ncolors - 3 C_F) \left( 1 + \frac{\pi^2}{12} \right) 
\bigg \rbrace
+ \Ncolors F_{gg}(z)
- \frac{1}{\epsilon_{\rm IR}} P_{gg}(z)   \Bigg] \Bigg \}
,
\label{eq:sigmapresub}
\eeq
where 
\beq
P_{gg}(z) = 2 \Ncolors \left[ \frac{1}{z} + \left( \frac{1}{1-z} \right) _+ + 
z(1-z) - 2 \right] + \frac{b_0}{2} \delta (1-z)
.
\label{eq:definePgg}
\eeq
Factorization requires the subtraction of 
\beq
\sigma_0 \frac{\alpha_S}{\pi} 
\frac{\mu^{4 \epsilon} \Gamma (1+\epsilon)}{1 - \epsilon}
 \left( \frac{4 \pi \mu^2}{Q_F^2} \right)^{\epsilon} 
\left( - \frac{1}{\epsilon_{\rm IR}} P_{gg}(z) + C_{gg}(z) \right) 
\eeq
from eq.~(\ref{eq:sigmapresub}), where $C_{gg}(z)$ is scheme-dependent. 
In the \MSbar factorization scheme that we will use for numerical
work below, $C_{gg}(z) = 0$, but it is a non-trivial function of $z$
in other schemes such as DIS.
Taking $\epsilon \rightarrow 0$ gives the final parton-level cross 
section
\beq
\hat{\sigma} (gg \rightarrow \eta_{\tilde{t}}+X) &=& 
\sigma_0 
\Bigg \{ \delta(1-z) + 
\frac{\alpha_S}{\pi} \Bigg[ 
\delta (1-z) \bigg\lbrace \frac{b_0}{2} \ln \left( \frac{Q^2}{M^2} \right) 
+(\Ncolors - 3 C_F)   \left( 1 + \frac{\pi^2}{12} \right) 
\bigg \rbrace
\nonumber \\ && 
- P_{gg} \ln 
\left( \frac{Q_F^2}{M^2} \right)+ \Ncolors F_{gg}(z)  - C_{gg}(z) 
\Bigg] \Bigg \} 
\label{eq:ggparton}.
\eeq
The form of this result is very similar to the corresponding one for toponium
production, eq.~(38) in ref.~\cite{Kuhn:1992qw}, 
when both are written in terms of their
respective LO contribution $\sigma_0$, differing only in the coefficient of 
$\sigma_0 C_F \alpha_S/\pi$.

\subsection{Quark-gluon scattering}

\begin{figure}[!tbp]
\begin{center}
\begin{picture}(80,80)(-40,-40)
\Photon(-45,-32)(0,-32){2}{8}
\ArrowLine(-45,32)(0,32)
\ArrowLine(0,32)(45,32)
\Photon(0,32)(0,0){2}{5}
\DashArrowLine(0,0)(45,0){4}
\DashArrowLine(0,-32)(0,0){4}
\DashArrowLine(45,-32)(0,-32){4}
\Text(42,10)[]{$\tilde{t}_1$}
\Text(42,-22)[]{$\tilde{t}_1^*$}
\Text(0,-46)[c]{(a)}
\end{picture}
~~~~~~~~~~~~~~~~~
\begin{picture}(80,80)(-40,-40)
\Photon(-45,-32)(0,-16){2}{8}
\ArrowLine(-45,32)(0,32)
\ArrowLine(0,32)(45,32)
\Photon(0,32)(0,-16){2}{8}
\DashArrowLine(0,-16)(45,0){4}
\DashArrowLine(45,-32)(0,-16){4}
\Text(42,10)[]{$\tilde{t}_1$}
\Text(42,-22)[]{$\tilde{t}_1^*$}
\Text(0,-46)[c]{(b)}
\end{picture}
\end{center}
\caption{\label{fig:quarkgluon}
The non-zero diagrams corresponding to the leading-order production of stop-antistop
in a color-singlet state through quark-gluon scattering. The additional diagram 
obtained from (a) by arrow reversal on the squark line
is included but not shown here.
}
\end{figure}
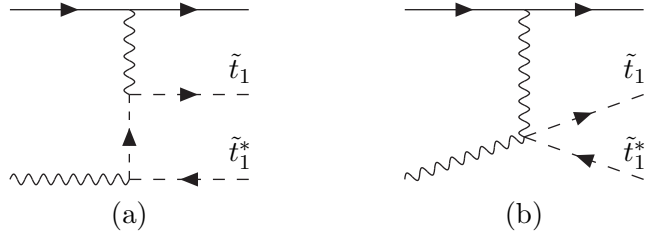

Quark-gluon scattering contributes at relative order $\alpha_S$
compared to the LO gluon fusion cross 
section. The diagrams for the process 
$qg \rightarrow q\tilde t \tilde t^*$
are given in Fig.~\ref{fig:quarkgluon}.%
Let $k_1$ be the initial-state gluon 
momentum and $k_2$ ($k_3$) be the initial (final) momentum of the 
approximately massless quark, 
and let the squark and anti-squark each have momentum $p$.
Then in terms of the 
parton-level Mandelstam variables $\hat{s}$, $\hat{t}$, and $\hat{u}$
defined in eq.~(\ref{eq:definestu}), 
the squared amplitude after summing over spin, 
polarization, and color is
\beq
\sum{|\mathcal{M}|^2} = \frac{4 g^6 C_F}{\hat t} \left [ 
\frac{2\hat{s} \hat{u}}{(\hat{t}-M^2)^2} - 1 + \epsilon \right ]
.
\label{eq:qgamp}
\eeq
Inserting the required factors to average over the initial spin,
polarization, and color, the parton-level differential cross section for
$qg \rightarrow \eta_{\tilde t} q$ is obtained by making the replacement
of eq.~(\ref{eq:dP3}) for $d\mathcal{P}_3$ in
\beq
d\hat{\sigma} (qg \rightarrow \eta_{\tilde{t}}q) = 
\frac{1}{2\hat{s}} \frac{1}{2 (d-2) \Ncolors (\Ncolors^2 - 1)} 
\sum{|\mathcal{M}|^2} d\mathcal{P}_3 
.
\eeq
The result is
\beq
d\hat{\sigma} (qg \rightarrow \eta_{\tilde{t}}q) & = & 
\sigma_0 \frac{\alpha_S}{2 \pi} C_F 
\frac{\mu^{4 \epsilon}}{\Gamma(2-\epsilon)} 
\left( \frac{4 \pi \mu^2 \hat{s}}{\hat{u} \hat{t}} \right)^{\epsilon} 
\frac{M^2}{\hat{s}\hat t} 
\left[ \frac{2\hat{s} \hat{u}}{(\hat{s}+\hat{u})^2} 
- 1+ \epsilon \right] d\hat{t}
.
\eeq
The angular integration required to obtain the parton cross section is 
performed in exactly the same way for gluon fusion.  Replacing the Mandelstam variables $\hat{s}$, $\hat{t}$, $\hat u$
with $y$ and $z$ using eqs.~(\ref{eq:definez}) and (\ref{eq:substitution}) 
results in
\beq
\hat{\sigma} (qg \rightarrow \eta_{\tilde{t}}q) & = & 
\sigma_0 \frac{\alpha_S}{2\pi}  
C_F \left( \frac{4 \pi \mu^2}{M^2} \right)^{\epsilon}  
\frac{\mu^{4 \epsilon}}{\Gamma(2-\epsilon)} z^{1+\epsilon} (1-z)^{-2\epsilon}
\nonumber \\ && 
\int_0^1 dy\> {y^{-\epsilon}(1-y)^{-1-\epsilon} \left ( \frac{1 
+ y^2(1-z)^2}{[1-y(1-z)]^2} - \epsilon \right )   }.
\eeq
Using the expansions in equations \eqref{eq:yplus} and \eqref{eq:zplus}, 
we integrate this to obtain
\beq
\hat{\sigma} (qg \rightarrow \eta_{\tilde{t}}q) & = & \sigma_0 
\frac{\alpha_S}{2\pi} 
\left( \frac{4 \pi \mu^2}{M^2} \right)^{\epsilon} 
\frac{\mu^{4 \epsilon}\Gamma(1+\epsilon)}{1 - \epsilon}
C_F
\Bigg [ \frac{1+(1-z)^2}{z} \left (-\frac{1}{\epsilon} 
+ 2 \ln(1-z) \right )
\nonumber \\ && 
 + 2 + z - 2/z - z \ln z \Bigg ].
 \label{eq:qgbeforesubtraction}
\eeq
The infrared divergence 
is removed in factorizing the cross section by subtracting 
\beq
\sigma_0 \frac{\alpha_S}{2\pi} 
\frac{\mu^{4 \epsilon}\Gamma(1+\epsilon)}{1 - \epsilon}
\left( \frac{4 \pi \mu^2}{Q_F^2} \right)^{\epsilon} 
\left( - \frac{1}{\epsilon} P_{gq}(z) + C_{gq}(z) \right) 
\eeq
from eq.~(\ref{eq:qgbeforesubtraction}),
where the splitting function $P_{gq}(z)$ is defined by
\beq
P_{gq}(z) = C_F [1 + (1-z)^2]/z
\eeq
and $C_{gq}(z) = 0$ in the $\overline{\rm MS}$ factorization scheme that
we will use for numerical work. Now taking $\epsilon \rightarrow 0$, we
arrive at the final result for the order-$\alpha_S^3$ quark-gluon parton
cross section
\beq
\hat{\sigma} (qg \rightarrow \eta_{\tilde{t}}q) =  
\sigma_0 \frac{\alpha_S}{2\pi}
\left [ P_{gq}(z) \ln 
\left ({M^2(1-z)^2}/{Q_F^2}\right )  
+ C_F (2 + z - 2/z -z \ln z) - C_{gq}(z)\right ].
\phantom{xxx}
\label{eq:Qgparton}
\eeq
Just as we found for the real gluon emission corrections to the leading order 
diagrams, this result is identical to the corresponding result for the 
production of quarkonium (eq.~(47) of ref.~\cite{Kuhn:1992qw}),
when both are written in terms of their respective LO results $\sigma_0$
for gluon fusion. The cross section for antiquark-gluon scattering 
$(\bar qg \rightarrow \eta_{\tilde{t}}\bar q)$ is also given by 
eq.~(\ref{eq:Qgparton}). 

\subsection{Quark-antiquark annihilation}

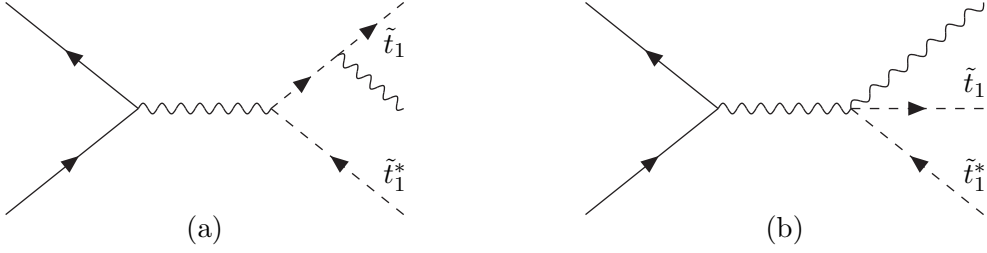
\begin{figure}[!tbp]
\begin{center}
\begin{picture}(150,80)(-75,-40)
\ArrowLine(-75,-40)(-25,0)
\ArrowLine(-25,0)(-75,40)
\Photon(-25,0)(25,0){2}{7}
\DashArrowLine(25,0)(50,20){4}
\DashArrowLine(50,20)(75,40){4}
\DashArrowLine(75,-40)(25,0){4}
\Photon(50,20)(75,0){2}{5}
\Text(72,25)[]{$\tilde{t}_1$}
\Text(72,-25)[]{$\tilde{t}_1^*$}
\Text(0,-46)[c]{(a)}
\end{picture}
~~~~~~~~~~~~~~~~~
\begin{picture}(150,80)(-75,-40)
\ArrowLine(-75,-40)(-25,0)
\ArrowLine(-25,0)(-75,40)
\Photon(-25,0)(25,0){2}{7}
\DashArrowLine(25,0)(75,0){4}
\DashArrowLine(75,-40)(25,0){4}
\Photon(25,0)(75,40){2}{7}
\Text(72,10)[]{$\tilde{t}_1$}
\Text(72,-25)[]{$\tilde{t}_1^*$}
\Text(0,-46)[c]{(b)}
\end{picture}
\end{center}
\caption{\label{fig:quarkantiquark}
Non-zero diagrams corresponding to the leading-order production of 
stop-antistop in a color-singlet state
through quark-antiquark annihilation. 
Note that these are related by crossing to the diagrams of figure 
\ref{fig:quarkgluon}.
The additional diagram 
obtained from (a) by arrow reversal on the squark line
is included but not shown here.}
\end{figure}

The production of color singlet stoponium through quark-antiquark annihilation is
impossible at leading order.  However, the process can occur through the emission of 
a real gluon, through the diagrams of figure \ref{fig:quarkantiquark},%
which are related to the diagrams for quark-gluon scattering in figure
\ref{fig:quarkgluon} by crossing 
the initial gluon line and the final quark line.  Therefore, the 
quark-antiquark annihilation squared
amplitude $\sum |\mathcal{M}|^2$ can be obtained directly from 
eq.~(\ref{eq:qgamp})
by the substitutions $(\hat s, \hat t, \hat u) \rightarrow 
(\hat u, \hat s, \hat t)$, 
with an extra  factor of $-1$ because we are crossing a fermion. 
This immediately yields 
\beq
\sum |\mathcal{M}|^2  = 
-\frac{4 g^6 C_F}{\hat s} \left [ 
\frac{2\hat{t} \hat{u}}{(\hat s - M^2)^2} - 1 + \epsilon \right ] 
.
\eeq
There is no possible infrared divergence, 
so we can immediately take $\epsilon = 0$. 
Using
\beq
d \hat \sigma (q\overline q \rightarrow \eta_{\tilde t} g) = 
\frac{1}{2 \hat s} \frac{1}{4 \Ncolors^2} \sum |\mathcal{M}|^2 d \mathcal{P}_3
\eeq
and applying the substitutions in eqs.~(\ref{eq:dP3}) and (\ref{eq:substitution}), we have 
\beq
\hat{\sigma} (q\bar{q} \rightarrow \eta_{\tilde{t}}g) 
= \sigma_0 \frac{\alpha_S}{\pi} C_F^2 z(1-z) \int_0^1 [1-2y(1-y)]\, dy
= \sigma_0 \frac{\alpha_S}{\pi} C_F^2 \frac{2}{3} z (1-z).
\label{eq:sigmaQQbar}
\eeq
This has the same form as the corresponding parton cross-section for
toponium (eq.~(49) of ref.~\cite{Kuhn:1992qw}), 
when both are written in terms of their
respective leading order gluon fusion results $\sigma_0$.

\section{Cross sections at hadron colliders\label{sec:hadron}}
\setcounter{footnote}{1}
\setcounter{equation}{0}

The parton model and QCD factorization 
relate the hadron-level\footnote{We will write formulas as they
would apply to $pp$ colliders such as the LHC, but simple changes 
can be applied in the obvious way to
obtain $p\bar p$ collider results.} cross section to the 
parton cross sections through:
\beq
\sigma (pp \rightarrow \eta_{\tilde{t}}+ X) = 
\sum_{a,b} 
\int
dx_a 
dx_b \  f_a^p(x_a, Q_F^2) f_b^p(x_b,Q_F^2)\,
\hat{\sigma} (ab \rightarrow  \eta_{\tilde{t}}+X) 
,
\eeq 
where $f_a^p(x_a, Q_F^2)$ is the parton distribution function 
for parton $a$ carrying momentum fraction $x_a$ in a proton 
probed at a scale $Q_F$.
Since the parton cross sections are integrated in terms of $z$, 
we make the change of variables
%
$x_a = x$, 
$x_b = \tau/xz$,
where
$\tau \equiv M^2/s$,
where $\sqrt{s}$ is the hadron-level center-of-momentum energy. 
This gives
\beq
\sigma (pp \rightarrow \eta_{\tilde t} + X) =
\sum_{a,b} \int_\tau^1 dz \int_{\tau/z}^1 dx \, \frac{\tau}{x z^2}
\, f_a^p(x, Q_F^2) f_b^p(\tau/xz, Q_F^2) 
\, 
\hat{\sigma} (ab \rightarrow  \eta_{\tilde{t}}+ X)
.
\label{eq:sigmappxz}
\eeq
[It is sometimes useful in numerical work to
decouple the limits of integration by making the further change of variables
$x = 1 - v(1 - \tau/z)$.]

Equation (\ref{eq:sigmappxz}) is applied using eqs.~(\ref{eq:ggparton}),
(\ref{eq:Qgparton}) for both quark-gluon and antiquark-gluon scattering, and (\ref{eq:sigmaQQbar}).
Since the lower limit of $z$ integration is $\tau$, 
the plus distributions in the parton-level 
cross sections must be shifted in order to be easily integrated.  Define the ``$\tau+$'' distribution as
\beq
\left( F(x) \right)_{\tau +} 
= \lim_{\beta \rightarrow 0} \Bigg [ 
F(x) \Theta( 1 - x - \beta ) - 
\delta(1 - x - \beta) \int_{\tau}^{1-\beta} F(y) \, dy \Bigg ],
\eeq
which is related to the plus distribution by
\beq
\left( F(x) \right)_{\tau +} = \left( F(x) \right)_+ + \lim_{\beta \rightarrow 0} \delta(1-x-\beta) \int_{0}^{\tau}{F(y) \ dy}
.
\eeq
Now the plus distributions occurring in the gluon-fusion 
cross sections [see eqs.~(\ref{eq:defineFgg}), (\ref{eq:definePgg}) and 
(\ref{eq:ggparton})] are replaced using
\beq
\left( \frac{1}{1-z} \right)_+ & = & 
\left( \frac{1}{1-z} \right)_{\tau +} + \ln(1-\tau) \delta(1-z)
, 
\\
\left( \frac{\ln (1-z)}{1-z} \right)_+ & = & 
\left( \frac{\ln (1-z)}{1-z} \right)_{\tau +} + 
\frac{1}{2} \ln^2 (1-\tau) \delta (1-z),
\eeq
and integration of terms containing $\tau +$ distributions can proceed using 
\beq
\int_{\tau}^1 \left( F(z) \right)_{\tau +} G(z) \ dz = 
\int_{\tau}^1 \Big[ G(z) - G(1) \Big] F(z) \ dz
.\eeq
Note that the simplification in equation (\ref{eq:plussimplify}) is not affected by 
changing the lower limit of integration over the plus distributions.

\section{Numerical results\label{sec:numerical}}
\setcounter{footnote}{1}
\setcounter{equation}{0}

In this section we present our numerical results for the stoponium cross section at 
next-to-leading order in QCD in $pp$ collisions at energies relevant for the LHC.  
(We note in passing that the QCD cross section times branching ratio 
for $p \bar p \rightarrow \eta_{\tilde t}+X \rightarrow \gamma\gamma+X$
at $\sqrt{s} = 1.96$ TeV is much less than 1 fb even for stoponium as light as 200 GeV,
so searches for $\eta_{\tilde t} \rightarrow \gamma\gamma$ at the Fermilab Tevatron
are presumably hopeless unless there is some very strong new non-QCD production mechanism.)

\subsection{Stoponium wavefunction effects\label{subsec:wavefunction}}

In the static color singlet model all of the 
nonperturbative information about the formation of the bound state is 
contained in the amplitude of the wavefunction at the origin 
(and its derivatives, for non-zero angular momentum states). 
In section \ref{sec:parton}, the parton-level cross sections were obtained
in terms of $|R(0)|^2/M^3$, a quantity that can be calculated approximately
by using potential models that simulate the effects of QCD. 
A naive Coulombic model for the QCD binding force would imply that
for the $S$-wave level $n$ state ($n=1$ is the ground state), 
\beq
\frac{|R_{nS}(0)|^2}{M(ns)^3} = \frac{4\alpha_S^3}{27 n^3} ,
\label{eq:coulomb}
\eeq 
but asymptotic freedom and other perturbative and non-perturbative QCD effects
make this approximation quite crude. 
We will instead obtain the value of the 
wavefunction at the origin by using 
a parameterization \cite{Hagiwara:1990sq} based on a potential 
extrapolated from the study of charmonium and bottomonium. 
The numerical results for $|R_{nS}(0)|^2/M(ns)^3$ in 
ref.~\cite{Hagiwara:1990sq} depend 
on the 4-flavor QCD scale $\Lambda^{(4)}_{\rm QCD}$ as an input.
For the range 
$\alpha_S^{(5)}(m_Z) = 0.1185 \pm 0.0025$, the relevant range is 
$\Lambda^{(4)}_{\rm QCD} \approx 300 \ \text{MeV} \pm 40 \ \text{MeV}$, and 
so we will use the $\Lambda^{(4)}_{\rm QCD}  = 300$ MeV parameterizations 
from ref.~\cite{Hagiwara:1990sq} in our numerical analysis.

While the largest stoponium production is for the $1S$ state, it 
can also be produced in excited states, and the signal 
will be enhanced by their decays either directly to 
two photons or to lower states which then decay to 
two photons \cite{Drees:1993uw}.  As mentioned 
in this reference, the production of higher angular momentum states is 
essentially of relative order $\alpha_S^2$ \cite{Hagiwara:1990sq}, so we 
do not consider their direct production as a correction.  However the 
contributions of higher $S$-wave states will be non-negligible.  Although 
their annihilation decay branching ratios 
will be the same as the ground state, they may 
also decay to $P$-wave bound states that cannot 
annihilate directly to a two-photon final state, but can decay to the
ground state or other $S$-wave states.
The total diphoton signal is therefore presumably bounded above by the sum over
the production cross sections for all $nS$ states, but it is unknown
how much the higher $n$ states will contribute to the diphoton signal.
However, since the phase space for $2S \rightarrow 1P$ 
decays is highly suppressed due to a very small mass 
difference, the diphoton decay branching ratios of the $1S$ and $2S$ states 
should be nearly 
identical. In our numerical results we will therefore conservatively
assume that the relevant production is due to these two states, and 
use in the results of section \ref{sec:parton} 
the effective wavefunction at the origin factor:
\beq
\frac{|R(0)|^2}{M^3} \rightarrow \sum_{n=1}^{2} \frac{|R_{nS}(0)|^2}{M(nS)^3} 
\eeq
neglecting the contributions from $n\geq 3$. 
The higher energy states may also 
contribute significantly, but with diphoton 
branching ratios suppressed by an unknown factor 
due to available decays to $P$-wave bound states that do not 
eventually
decay to diphotons. 
Furthermore, it is quite possible that the potential 
models become less reliable for the higher excited states.

\begin{figure}[!tbp]
\begin{minipage}[]{0.47\linewidth}
\caption{\label{fig:wavefunction}
The bound state wavefunction at the origin factor $\sum |R(0)|^2/M^3$,
as a function of the 1S bound 
state mass, as parameterized in \cite{Hagiwara:1990sq}.  The lower three lines are for the 
1S wavefunction only, the middle group is the sum of the 1S and 2S wavefunctions, 
and the top group is the sum of the lowest 10 $S$-wave bound state wavefunctions.
The solid lines are the $\Lambda^{(4)}_{\rm QCD} = 300$ MeV parameterization and the 
dashed lines show the $\pm$40 MeV variations in this value.}
\end{minipage}
\begin{minipage}[]{0.52\linewidth}
\includegraphics[width=7.5cm,angle=0]{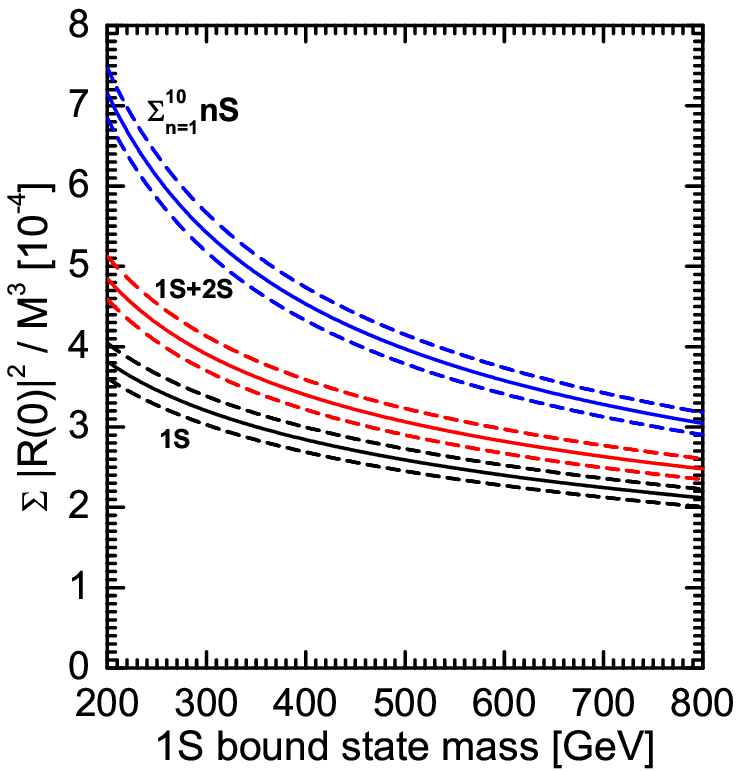}
\end{minipage}
\end{figure}
In Figure \ref{fig:wavefunction} we plot the effective wavefunction at 
the origin factor given in \cite{Hagiwara:1990sq} 
as a function of the 1S bound state mass.  Three cases are 
shown in this plot: the contribution from the 1S bound state alone, the 
sum of the 1S and 2S bound states that we will use in numerical work, 
and the sum of the first 10 $S$-wave 
states.  In each case, the 40 MeV variation 
in the QCD scale 
$\Lambda^{(4)}_{\rm QCD}$ (obtained by interpolation from the
values given in ref.~\cite{Hagiwara:1990sq}) is also plotted on the graph.
Although the probability for producing higher-energy bound states drops 
sharply with higher $n$, 
their sum can add up to a significant enhancement to the overall 
production rate.\footnote{It is interesting to note that the results for the wavefunction
at the origin obtained in \cite{Hagiwara:1990sq} are considerably smaller
than would be obtained from the naive Coulombic formula (\ref{eq:coulomb})
using $\alpha_S(Q)$ evaluated at $Q = 1/\langle r_{nS} \rangle =
2 \alpha_S(Q) M_{\eta_{\tilde t}}/9n^2$. However, the results of 
eq.~(\ref{eq:coulomb}) depend very sensitively on this somewhat arbitrary
choice of scale.
Also, the contribution from higher $n$ states from \cite{Hagiwara:1990sq}
is larger relative to the ground state contributions than is suggested by
the Coulombic formula.} 
It is not unlikely that we have
underestimated the cross section for stoponium production and the rate of 
diphoton annihilation decays, but precise calculation of the additional 
enhancements will have to await a better understanding of stoponium
spectroscopy.

\begin{figure}[!tbp]
\begin{minipage}[]{0.47\linewidth}
\caption{\label{fig:bindingenergy}
Binding energies $2m_{\tilde t} - M_{\eta_{\tilde t}}(nS)$ of the nS 
bound states of mass $M_{\eta_{\tilde t}}(nS)$ as a function of the 
constituent squark mass $m_{\tilde t}$.  The top line is the 1S state, 
the second is the 2S state, and so on to the 10S state at the bottom.  
This plot uses the $\Lambda^{(4)}_{\rm QCD} = 300$ MeV parameterization in 
ref.~\cite{Hagiwara:1990sq}.}
\end{minipage}
\begin{minipage}[]{0.52\linewidth}
\includegraphics[width=7.5cm,angle=0]{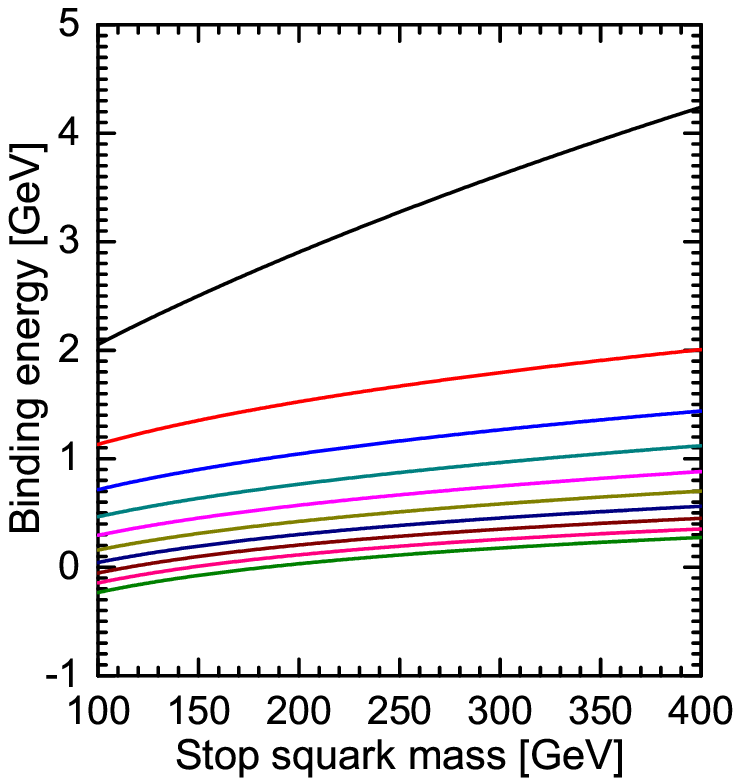}
\end{minipage}
\end{figure}

In Figure \ref{fig:bindingenergy}
we plot the binding energies of the 
first 10 $S$-wave bound states, computed using the parameterization in 
\cite{Hagiwara:1990sq}.  These are about 1\% of the 1S bound state mass 
and significantly less for any other bound state.  These differences are 
within the scale dependence of our result, and more importantly they will 
be below the energy resolution of the detectors. Therefore we will ignore 
the small binding energies in the calculation of the parton cross 
sections, taking $M = 2m_{\tilde t}$ everywhere and ignoring any 
small differences in excited state masses.

\subsection{Numerical results for stoponium production}

To integrate numerically the parton-level cross sections in equations
\eqref{eq:ggparton}, \eqref{eq:Qgparton}, and \eqref{eq:sigmaQQbar} using eq.~(\ref{eq:sigmappxz}), we
will use the Martin-Stirling-Thorne-Watt (MSTW) 2008 NLO parton
distribution functions \cite{Martin:2009iq}, using their global best fits. 
These PDFs use the \MSbar subtraction scheme, and so we set 
$C_{gg}(z) =0$ in eq.~(\ref{eq:ggparton}) and $C_{gq}(z) = 0$ in
eq.~(\ref{eq:Qgparton}). To be consistent with MSTW \cite{Martin:2009bu}, 
we run $\alpha_S(Q)$ with
two-loop beta function in the $n_f = 5$ effective theory, 
starting from $\alpha_S(m_Z) = 0.12018$.

\begin{figure}[!tbp]
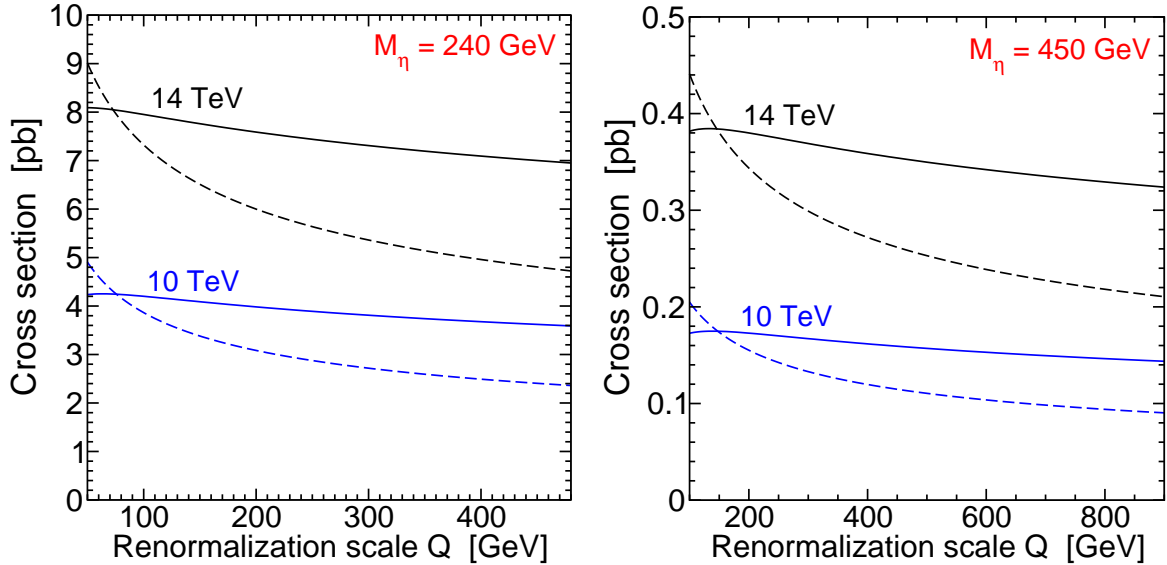

\includegraphics[width=7.5cm,angle=0]{varQ_240.eps}~~
\includegraphics[width=7.5cm,angle=0]{varQ_450.eps}
\caption{\label{fig:scaledependence}
The scale dependence of the LO (dashed) and NLO (solid) 
$pp \rightarrow \eta_{\tilde t}+X$
cross sections for $M_{\eta_{\tilde t}} = 240$ GeV (left) and 450 GeV (right), 
with the factorization scale $Q_F$ set equal to the renormalization scale $Q$,
for $\sqrt{s} = 14$ and 10 TeV.}
\end{figure}
In Fig.\ \ref{fig:scaledependence} we plot the renormalization scale 
dependence of the NLO $pp \rightarrow \eta_{\tilde t}+X$ 
cross section for two different masses $M_{\eta_{\tilde t}} = 240$ and 450 GeV, 
varying $50\,{\rm GeV} < Q < 480\,{\rm GeV}$ and
$100\,{\rm GeV} < Q < 900\,{\rm GeV}$ respectively as 
reasonable ranges for the choice of the common renormalization and factorization 
scale $Q = Q_F$. The figure shows the results for
both $\sqrt{s} = 10$ TeV and 14 TeV proton-proton colliders.
These results show a significantly improved renormalization scale dependence 
for the NLO result compared to the LO cross section in each case.

\begin{figure}[!tbp]
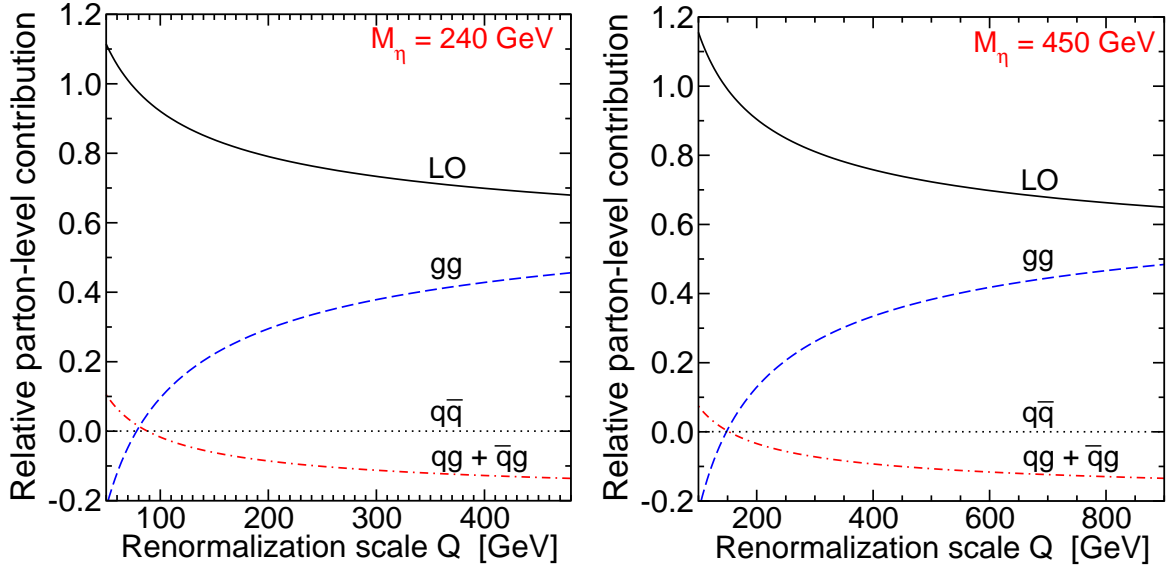

\includegraphics[width=7.5cm,angle=0]{varQ_240_rel.eps}~~
\includegraphics[width=7.5cm,angle=0]{varQ_450_rel.eps}
\caption{\label{fig:relparton}
Relative contributions from each parton-level process relative to the 
total NLO $pp \rightarrow \eta_{\tilde t}+X$ for 
$M_{\eta_{\tilde t}}$ = 240 GeV (left) and 450 GeV (right), for
$\sqrt{s} = 14$ TeV, as a function of varying $Q = Q_F$,
as in figure \ref{fig:scaledependence}.  
The LO contribution is the order $\alpha_S^2$ result, 
the $gg$ contribution is due to real and virtual corrections to the 
leading order diagrams (the remaining part of eq.~(\ref{eq:ggparton}), 
the $qg + \bar{q} g$ contribution is from quark-gluon and antiquark-gluon
scattering eq.~(\ref{eq:Qgparton}), 
and the $q\bar{q}$ contribution is from quark-antiquark annihilation
eq.~(\ref{eq:sigmaQQbar}).}
\end{figure}

We present the individual
parton-induced contributions relative to the full hadronic cross section in 
Fig.~\ref{fig:relparton},
for $\sqrt{s} = 14$ TeV 
proton-proton colliders and for the same masses and over the same
ranges of $Q$ as in figure \ref{fig:scaledependence}.%
Not surprisingly, for $Q = Q_F = M_{\eta_{\tilde t}}$, the
relative contributions from each parton-level process are quite similar to
those found for the toponium cross section in ref.~\cite{Kuhn:1992qw}. 
It is also interesting to note that for the scale choice 
$Q = Q_F \approx M_{\eta_{\tilde t}}/3$, the NLO $gg$, $qg+\bar q g$ and $q\bar q$
corrections are all simultaneously small. We have checked that this also holds for
lower beam energies.

\begin{figure}[!tbp]
\begin{minipage}[]{0.52\linewidth}
\caption{\label{fig:NLOcross}
Results for NLO order cross sections for 
$pp \rightarrow \eta_{\tilde t}+X$
with $M_{\eta_{\tilde t}}$ between 200 and 800 GeV.  
Here we have set the factorization and renormalization scales equal to the stoponium mass, $Q = Q_F = M_{\eta_{\tilde t}}$.}
\end{minipage}
\begin{minipage}[]{0.47\linewidth}
\begin{flushright}
\includegraphics[width=7.5cm,angle=0]{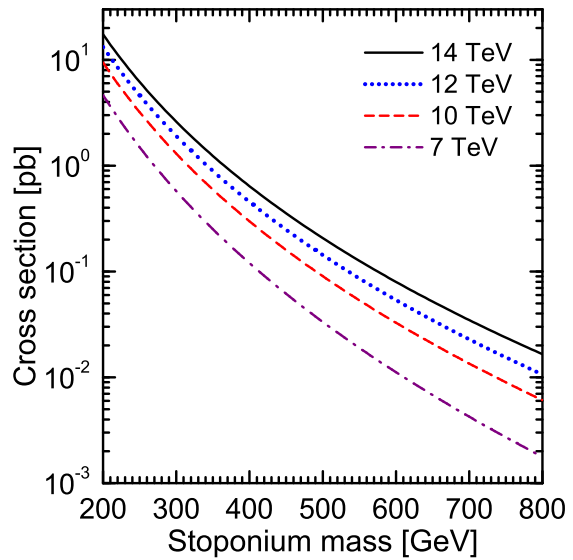}
\end{flushright}
\end{minipage}
\end{figure}

In Fig.~\ref{fig:NLOcross},%
we show the NLO $pp \rightarrow \eta_{\tilde t} + X$ cross section as a
function of $M_{\eta_{\tilde t}}$, with the renormalization and factorization
scales set equal to the stoponium mass. At this writing, the ultimate
performance of the LHC is the subject of considerable speculation, so we
show results for four different beam energies. 
\begin{figure}[!tbp]
\begin{minipage}[]{0.48\linewidth}
\caption{\label{fig:NLOK}
The K-factor, defined as the ratio of the NLO to LO cross sections for
$pp \rightarrow \eta_{\tilde t}+X$, as a function of the stoponium mass 
$M_{\eta_{\tilde t}}$, computed with $Q = Q_F = M_{\eta_{\tilde t}}$. 
Results are shown for $\sqrt{s} = 14$,
12, 10, and 7 TeV.}
\end{minipage}
\begin{minipage}[]{0.51\linewidth}
\begin{flushright}
\includegraphics[width=7.6cm,angle=0]{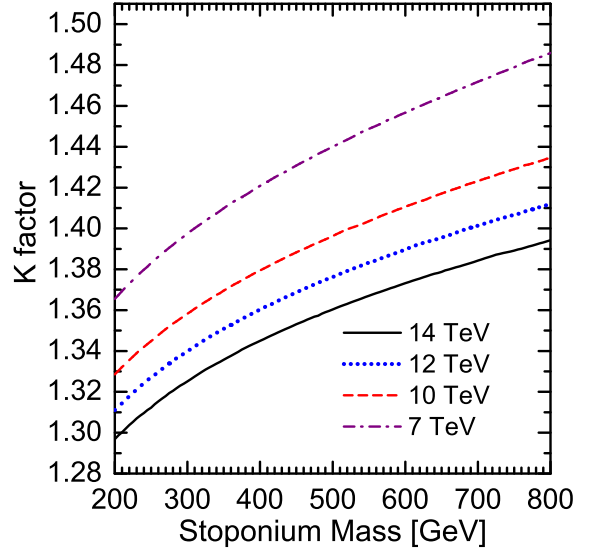}
\end{flushright}
\end{minipage}
\end{figure}
Figure \ref{fig:NLOK} shows the corresponding K-factor, defined as
the ratio of the next-to-leading order to the leading-order cross section,
again with $Q = Q_F = M_{\eta_{\tilde t}}$. Our results show that the enhancement
in the stoponium production cross section due to the NLO corrections with this scale choice is
between 30\% and 50\%, depending on the energy of LHC, with larger
enhancements for larger $M_{\eta_{\tilde t}}$ and for smaller $\sqrt{s}$. 

\subsection{Numerical results for the diphoton final state}

\begin{figure}[!tbp]
\begin{minipage}[]{0.47\linewidth}
\caption{\label{fig:diphotonBR}
The NLO branching ratio of stoponium to two photons in the 
idealized approximation 
that the hadronic partial width dominates the full width, 
from eq.~(\ref{eq:Roneresult}).  We have used 
$\alpha_S(m_Z) = 0.12018$ and set the renormalization scale 
equal to the stoponium mass.}
\end{minipage}
\begin{minipage}[]{0.52\linewidth}
\includegraphics[width=7.5cm,angle=0]{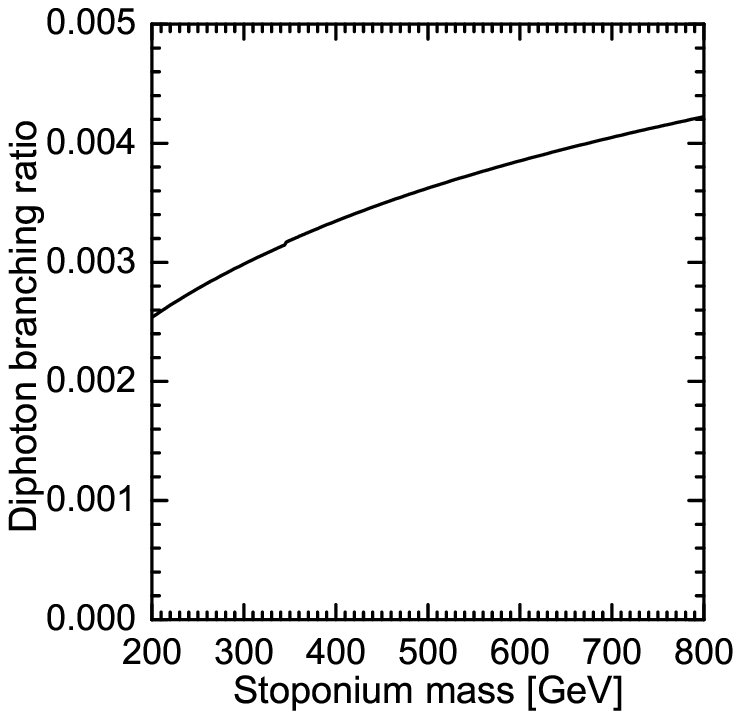}
\end{minipage}
\end{figure}

The observable signal for stoponium production is the diphoton peak
produced by the decay 
$\eta_{\tilde t} \rightarrow \gamma \gamma$.
The cross section
for this signal is the product of the production cross section we have calculated above and the model-dependent branching ratio for diphoton decays of stoponium:
\beq
\sigma(pp \rightarrow \eta_{\tilde t} \rightarrow \gamma \gamma) = 
\sigma(pp \rightarrow \eta_{\tilde t})
{\rm BR}(\eta_{\tilde t} \rightarrow \gamma \gamma) 
\eeq
Unfortunately, the $\gamma \gamma$ branching ratio
is in general highly dependent on the parameter space of soft supersymmetry 
breaking. However, one can proceed by first considering the idealized
model-independent case in which the total width is dominated by the gluon-gluon partial width. Then 
we can approximate the full width by the NLO width 
to hadrons ($gg+X$ or $q\bar{q} + X$), and calculate an
approximate branching ratio:
\beq
{\rm BR}(\eta_{\tilde t} \rightarrow \gamma \gamma) \approx
\frac{\Gamma^{(1)}(\eta_{\tilde{t}} \rightarrow 
\gamma \gamma)}{\Gamma^{(1)}(\eta_{\tilde{t}} \rightarrow 
\mbox{hadrons})} 
\equiv 
R^{(1)},
\eeq
where $\Gamma^{(1)}(\eta_{\tilde{t}} \rightarrow \mbox{hadrons})$ and $\Gamma^{(1)}(\eta_{\tilde{t}} \rightarrow 
\gamma \gamma)$ are the NLO partial widths.  
In our earlier paper \cite{Martin:2009dj}, we found 
\beq
R^{(1)} & = & 
\frac{8\alpha^2}{9\alpha_S^2} 
\bigg\{ 1 + \frac{\alpha_S}{\pi} \bigg[ -\frac{b_0}{2} 
\ln \biggl( \frac{Q^2}{M^2} \biggr) + 
\left(\frac{13 \pi^2}{24}-\frac{199}{18} \right) \Ncolors  
 + \left(\frac{\pi^2}{4} -2 - 2 \lntwo \right) C_F \nonumber \\
 & & + \left( \frac{8}{9}(n_f + n_t) 
+ n_t h(4 m_t^2 / M^2) 
+ \frac{1}{6} \lntwo \right) \bigg] \bigg\},
\label{eq:Roneresult}
\eeq
in which $M$ is the bound state mass, $n_f = 5$, 
$n_t$ is zero or one depending on whether or not the mass of stoponium 
is large enough for top-antitop decays to be kinematically allowed, 
and the function $h(r)$ is defined by
\beq
h ( r ) = \frac{2}{9} (4 - r) \sqrt{1-r} - \frac{8}{9}  
- \frac{2}{3} \ln(1+\sqrt{1-r})
+ \frac{2}{3} \lntwo. 
\label{eq:h(r)}
\eeq
The resulting idealized diphoton branching ratio
$R^{(1)}$ is shown in Fig.~\ref{fig:diphotonBR}.
(Here we have differed slightly from figure 8 of ref.~\cite{Martin:2009dj}
by using values of $\alpha_S(Q)$ that follow from $\alpha_S(m_Z) = 0.12018$
to agree with those used in MSTW \cite{Martin:2009bu},
using a slightly different two-loop
running that includes the stop squark and includes the top quark only if
the stop mass is greater than the top mass.  We
have also fixed the renormalization scale equal to the stoponium mass; the 
NLO scale variation is quite small as shown in \cite{Martin:2009dj}.)

\begin{figure}[!tbp]
\begin{minipage}[]{0.47\linewidth}
\caption{\label{fig:NLOdiphoton}
The NLO cross section times branching ratio for
$pp \rightarrow \eta_{\tilde t} +X$ followed by $\eta_{\tilde t} \rightarrow \gamma\gamma$,
in the idealized case that the stoponium decay width is dominated by the
decays to gluons. 
The factorization and renormalization scales are set equal to the stoponium mass.}
\end{minipage}
\begin{minipage}[]{0.52\linewidth}
\includegraphics[width=7.5cm,angle=0]{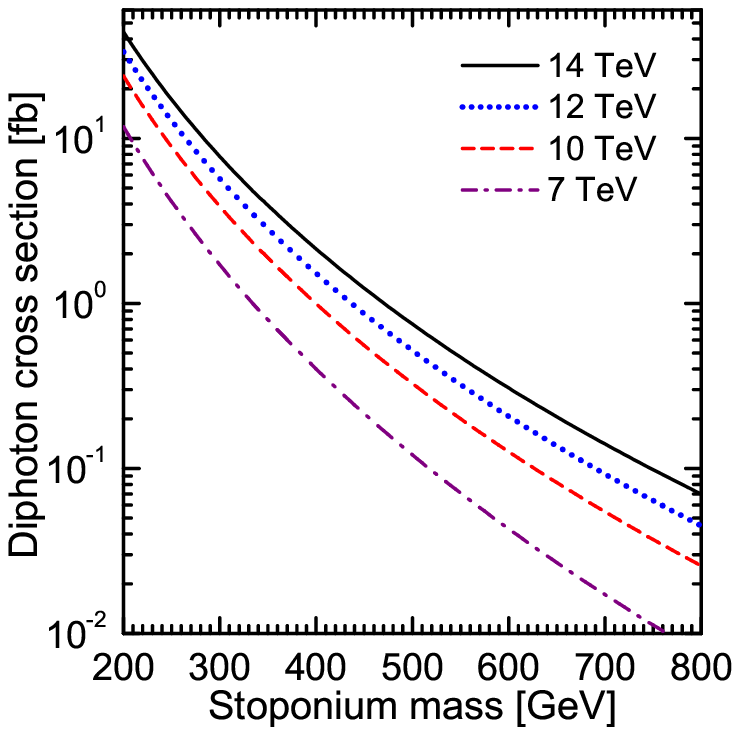}
\end{minipage}
\end{figure}

Now combining the NLO production cross section 
of fig.~\ref{fig:NLOcross} with the NLO branching ratio of 
fig.~\ref{fig:diphotonBR}, we find the
idealized NLO $pp \rightarrow \eta_{\tilde t} +X \rightarrow 
\gamma\gamma +X$ cross section shown in fig.~\ref{fig:NLOdiphoton}.
In this paper, we will not attempt a new study of the diphoton signal viability 
including backgrounds and after cuts.
However, it is interesting to compare the corrected results 
we have found in the present paper
and in \cite{Martin:2009dj} to the corresponding LO results that were used as the
basis for the study in ref.~\cite{Martin:2008sv}. 
There are several new effects taken into
account in this paper. First, the NLO corrections to the diphoton branching ratio
lead to a decrease of 30\% to 35\% when evaluated at $Q = M_{\eta_{\tilde t}}$. 
Second, and counteracting this, 
the NLO corrections to the production cross section lead to an increase
as shown in fig.~\ref{fig:NLOK}. 
Third, we are including here the effects of the $2S$ state as discussed in section 
\ref{subsec:wavefunction}. Also, ref.~\cite{Martin:2008sv} used a different set of
parton distribution functions and a different $\alpha_S$.
The net result of these effects is shown in Table \ref{tab:compareold}, which compares
the cross section (before cuts) used in the study ref.~\cite{Martin:2008sv}
to that found here.
\begin{table}[!tbph]
\centering
\begin{tabular}{|c|c|c|c|}
\hline
$M_{\eta_{\tilde t}}$ [GeV] & $\sigma\times \rm{BR}$ (fb, NLO)
& 
$\sigma\times \rm{BR}$ (fb, LO in ref.~\cite{Martin:2008sv}) & 
\ ratio \ \\
\hline
\hline 
235 & 22.4 & 23.1 & 0.969 \\     
\hline
240 & 20.5 & 21.0 & 0.974 \\     
\hline
255 & 15.8 & 16.0 & 0.989 \\     
\hline
270 & 12.3 & 12.3 & 1.00 \\     
\hline
400 & 2.14 & 1.94 & 1.10 \\     
\hline
450 & 1.24 & 1.09 & 1.13 \\     
\hline
500 & 0.750 & 0.647 & 1.16 \\    
\hline
600 & 0.307 & 0.255 & 1.20 \\    
\hline
700 & 0.141 & 0.113 & 1.24 \\    
\hline
800 & 0.0700 & 0.0549 & 1.27 \\    
\hline
\end{tabular}
\caption{\label{tab:compareold}
Comparison of cross section times branching ratio for $pp \rightarrow 
\eta_{\tilde t} + X \rightarrow \gamma\gamma + X$
in proton-proton collisions at $\sqrt{s} = 14$ TeV, before cuts
and in the idealized case that the hadronic partial width dominates the
decay width of stoponium. The second column shows the results in 
fig.~\ref{fig:NLOdiphoton}, and
the third column shows the LO results as used in the study of
the signal including cuts and LO backgrounds in ref.~\cite{Martin:2008sv}.}
\end{table}
The results found here for the cross section times branching ratio are about
equal to those used in \cite{Martin:2008sv} for 235 GeV $< M_{\eta_{\tilde t}} <$ 270
GeV relevant for electroweak-scale baryogenesis models, and are somewhat larger for larger $M_{\eta_{\tilde t}} > 400$ relevant for compressed supersymmetry models.

In the two motivated scenarios mentioned in the Introduction, there are
somewhat different expectations for the reduction
of the $pp\rightarrow \eta_{\tilde t}+X \rightarrow \gamma\gamma + X$
cross section compared to the idealized case just mentioned.
For compressed supersymmetry \cite{compressed}, 
in which enhancements to dark matter annihilation are mediated by $t$-channel stop squark exchange, 
the ground state stoponium will have a mass between about 400 GeV and 800 GeV,
tending towards the former in models with less fine-tuning. The stoponium
total width is indeed dominated by the hadronic partial width, which has a
branching ratio that is typically more than 80\% and usually of order 90\%
or even higher \cite{Martin:2008sv}. 
The resulting diphoton 
cross sections can therefore be estimated from figure 
\ref{fig:NLOdiphoton}
simply by applying this correction factor as appropriate.

It is somewhat harder to make a prediction for the diphoton signal in models of 
electroweak-scale
baryogenesis \cite{EWbaryogenesis,EWbaryogenesisnew},
because in those models the $h^0h^0$ decay width typically dominates 
if it is
kinematically allowed, and also because the $WW$ and $ZZ$ modes are 
more significant.
We have 
calculated the radiative corrections to $h^0h^0$ decay channel in 
ref.~\cite{Martin:2009dj}, but the weak vector boson
decays are also significant (contributing up to about 30\% of the total
width for some choices of parameters). But to make a rough general statement
about the cross section for diphoton annihilations, we can 
assume that the corrections to the weak decay channels should probably not be
much larger than those for the hadronic decays; then the 30\% reduction in
$R^{(1)}$ over the leading order ratio leads us to conclude that a
corresponding reduction in 
${\rm BR}(\eta_{\tilde t} \rightarrow \gamma\gamma)$ is reasonable.  
These assumptions and the results in refs.~\cite{Martin:2008sv,Martin:2009dj}
allow us to make a conservative
estimate for the next-to-leading order branching ratio in the electroweak
baryogenesis models:
\beq
{\rm BR}(\eta_{\tilde t} \rightarrow \gamma \gamma) \,\gsim\,
\Biggl
\{ \begin{array}{l} 
1.5 \times 10^{-3}
\qquad (M < 2m_{h^0}),
\\
5.3 \times 10^{-4}
\qquad (M > 2m_{h^0}), 
\end{array}
\Biggr.
\label{eq:EWBGestimates}
\eeq
Note that in these models, one expects $M_{\eta_{\tilde t}}$ between about
235 and 270 GeV, which happens to be in just the range where the 
decays to $h^0$ might become kinematically allowed. 
The branching ratio is highly dependent on the Higgs scalar boson 
mass because of its
large partial width immediately above the threshold for Higgs pair
production.  Our resulting estimates for the stoponium diphoton
annihilation cross section in the Electroweak Baryogenesis scenario are
given in Table \ref{table:EWBGcross}.
\begin{table}[!tbph]
\centering
\begin{tabular}{|c|c|c|c|}
\hline
$M_{\eta_{\tilde t}}$ [GeV] & $\sigma(pp \rightarrow \eta_{\tilde t})$
 [pb] & $\sigma_{\gamma \gamma}$, $M<2m_{h^0}$ [fb] &
$\sigma_{\gamma \gamma}$, $M>2m_{h^0}$ [fb] \\
\hline
\hline
235 & 8.25 & 12 & 4.4 \\
\hline
255 & 5.63 & 8.4 & 3.0 \\
\hline
270 & 4.30 & 6.5 & 2.8 \\
\hline
\end{tabular}
\caption{\label{table:EWBGcross} Estimates of the NLO cross section for
the production and subsequent diphoton annihilation decay of stoponium in
a $\sqrt{s} = 14$ TeV $pp$ collider, as a function of the stoponium mass
in the electroweak baryogenesis model scenario. We have used approximate
$\gamma\gamma$ branching ratios from eqs.~(\ref{eq:EWBGestimates}).}
\end{table}

\section{Outlook\label{sec:outlook}}
\setcounter{footnote}{1}
\setcounter{equation}{0}

In this paper, we have derived the NLO QCD corrections to the cross section for
stoponium productions in
hadronic collisions relevant for the LHC. 
These could be important in understanding an eventual discovery of, or limits on,
stoponium through its diphoton annihilation decays.
We found that when calculated
at a scale $Q = M_{\eta_{\tilde t}}$, the corrections to the cross section are large and positive, partly
canceling the large negative corrections in the diphoton branching ratio
found in ref.~\cite{Martin:2009dj}. We also found that the NLO corrections to the
cross section are small
when evaluated at $Q \approx M_{\eta_{\tilde t}}/3$.
For stoponium of mass 240 GeV (450 GeV), we found that reducing the center of mass
energy from 14 TeV to 12 TeV will reduce the cross section by about 25\% (30\%),
while reducing from 14 to 10 TeV will reduce the cross section by about
52\% (55\%), with of course larger reductions for heavier masses. We have not attempted here a detailed study improving on the 
estimate of the reach of the LHC based on LO signal and background
in \cite{Martin:2008sv}. One prerequisite for such an improvement
is a better understanding of the Standard Model and detector diphoton backgrounds. 
There has been considerable study of the Standard Model high energy diphoton production at the LHC
\cite{Balazs:1999yf}-\cite{Balazs:2006cc}, but focused on lower invariant masses
relevant to a light Higgs scalar boson search. Fortunately, in the future 
the LHC will provide its
own background estimate in the form of a sideband analysis, and this should be
at least as robust as any calculation can be, at least for the purposes of
bump-hunting.

The largest uncertainty in estimates of the stoponium cross section is clearly our
lack of detailed understanding of stoponium energy levels and bound state matrix
elements.
In this paper, we used the parameterization of \cite{Hagiwara:1990sq} 
for the wavefunction at the 
origin, but it must be recognized
that this was obtained in part by a significant extrapolation from charmonium and bottomonium
data. Even at the energy scales relevant for stoponium, the QCD binding potential
is far from the semi-classical Coulomb form, and more work is needed to understand the relevant matrix
elements. 
We have tried to remain relatively conservative by not including the possible effects
from $S$-wave resonance production at the $3S$ level and beyond. An eventual discovery of stoponium in the diphoton channel will
not only provide an accurate determination of the top squark mass, but an opportunity
to use data to learn more about a QCD bound-state system in an energy range where calculations
are under relatively better control than the known charmonium and bottomonium
states.

\bigskip \noindent
{\it Acknowledgments:} We are grateful to Tim Tait for helpful conversation.
This work was supported in part by the National Science Foundation grant 
number PHY-0757325.


\end{document}